\documentclass[preprints,article,accept,moreauthors]{mdpi} 

\firstpage{1} 
\pubvolume{1}
\issuenum{1}
\articlenumber{0}
\pubyear{2022}
\copyrightyear{2022}
\datereceived{} 
\dateaccepted{} 
\datepublished{} 
\hreflink{https://doi.org/} 

\Title{Enigmatic emission structure  around  Mrk 783: cross-ionization of a companion in 100 kpc away.}

\TitleCitation{Enigmatic emission structure  around Mrk 783}

\newcommand{\SII}{[S~{\sc ii}]}
\newcommand{\OIII}{[O~{\sc iii}]}
\newcommand{\NII}{[N~{\sc ii}]}
\newcommand{\HII}{H~{\sc ii}}

\newcommand{\HeII}{He\,{\sc ii}}
\newcommand{\HI}{H~{\sc i}}
\newcommand{\Ha}{H$\alpha$}

\newcommand{\Hb}{H$\beta$}
\newcommand{\kms}{\,\mbox{km}\,\mbox{s}^{-1}}
\newcommand{\ergs}{\,\mbox{erg}\,\mbox{s}^{-1}}
\newcommand{\SIIHa}{[S~{\sc ii}]/H$\alpha$}
\newcommand{\NIIHa}{[N~{\sc ii}]/H$\alpha$}
\newcommand{\OIIIHb}{[O~{\sc iii}]/H$\beta$}

\newcommand{\apj}{ApJ}
\newcommand{\apjl}{ApJL}
\newcommand{\aap}{A\&A}
\newcommand{\araa}{AnRevA\&A}

\newcommand{\aj}{AJ}
\newcommand{\mnras}{MNRAS} 

\newcommand{\pasp}{PASP}

\Author{Alexei V. Moiseev $^{1,2, \dagger,\ddagger}$*\orcidA{} and Aleksandrina A.  Smirnova $^{1,\ddagger}$\orcidB{} 
and Tigran A. Movsessian $^{3,\ddagger}$}

\AuthorNames{Alexei V. Moiseev  et al.}

\AuthorCitation{Moiseev  et a.}

\address{%
$^{1}$ \quad Special Astrophysical Observatory, Russian Academy of Sciences, Nizhny Arkhyz 369167, Russia; moisav@gmail.com\\
$^2$ \quad Sternberg Astronomical Institute, Moscow State University, Moscow, 119234 Russia
\\
$^3$ \quad Byurakan Astrophysical Observatory, 0213 Aragatsotn prov., Armenia
}

\corres{Correspondence: moisav@gmail.com;  (A.V.)}

\abstract{Mrk 783 is  a narrow-line Seyfert 1 galaxy that  possesses a relatively large two-sided radio emission extending up to 14 kpc from the active nucleus possibly connected with a large-scale ionized gas emission.  We obtained a deep \OIII{} image that revealed an extended  system of emission knots and diffuse ionized gas surrounding the main galaxy. The high-excited gas is related not only with the  radio structure, but also with tidal features illuminated by the active nucleus radiation up to the projected distance 41 kpc as it follows from the emission lines intensities and kinematics derived from the long-slit spectroscopic data. Moreover the part of the disk of the companion galaxy SDSS J130257.20+162537.1 located at 99 kpc projected distances to the north of Mrk 783  also falls in the AGN ionizing cone. It is {possible}  that Mrk~783 can be considered   as     `Hanny’s Voorwerp precursor', i.e. a galaxy that demonstrates  signs of sequential  switching from radio-loud to radio-quiet nuclear activity, in the moment before   falling of its ionization luminosity.
}
\begin{document}

\section{Introduction} 
\label{sec:intro}
Active galactic nucleus (AGN) feedback is important in the coevolution between AGN and its host galaxy. This fundamental physical processes  has an impact on the interstellar medium and the intergalactic environment, but it  has been under debate for a long time and is still not fully understood. Mrk 783 is an interesting example of a galaxy in which we can trace AGN radiation feedback on large  spatial scales outside the  host galaxy.

Mrk 783 was discovered by  \citet{Markarian1976},  its optical spectra   was first define as Narrow Line Seyfert 1 galaxy (NLS1) by \citet{OP1985}.  According to the current point of view, NLS1 is AGN with a relatively low mass of the central  black hole (10$^6$–10$^8$~M$_{\odot}$)  in an early stage of AGN evolution  \citep{Mathur2000MNRAS.314L..17M}.  Mrk 783 black hole mass M$_{BH}=4.3\cdot10^7$ M$_{\odot}$ is in the higher end of NLS1s distribution, its Eddington ratio is quite low (0.11)  and its \OIII{} line is quite strong with respect to H$_\beta$  \citep{Berton2015}.

In the radio band Mrk 783 was extensively investigated by \citet{Congiu2017a,Congiu2020}. They have found in the center of galaxy a  compact core  with a pc-scale jet as well as two-sided extended component (up to 14 kpc from the nucleus). The authors stressed, that the galaxy \textit{`is one of the few NLS1 showing such an extended emission at $z < 0.1$'.}  The small-scale jet and the large-scale radio emission are not aligned. Based on these facts and on the very steep spectral indexes   \citet{Congiu2017a}  concluded that the extended emission observed in Mrk~783  might be a relic and that the radio source might be in a quiescent period of its activity cycle. 

No less interesting Mrk 783 is in the optical range.  V-band image reveals low surface brightness extended structures  on the both sides of the galaxy nucleus that looks like tidal tail observed in interacting galaxies. The isophotes of the internal part of the galaxy indicate the presence of a second point-like structure  which might be the nucleus of the second galaxy involved in the proposed merging  \citep{Congiu2020}. Optical emission of the ionized gas is far more extended with respect to the radio emission  mostly  on the south-east side of the nucleus. The \OIII, H$_{\beta}$ and H$_{\alpha}$ lines were tracked up to $\sim$ 35 kpc from the nucleus in the spectrum obtained at the 6.5m  Magellan telescope, that makes this EELR (Extended Emission Line Region) one of the most extended discovered so far \citep{Congiu2017b}. 

In order to better understand the physics of processes occurring in Mrk 783, we have mapped this galaxy in the \OIII{} emission line  at the 2.5m telescope of the Caucasus Mountain Observatory (CMO) of Sternberg Astronomical Institute of Moscow State University (SAI MSU) and also observed it at the 6m telescope of the Special Astrophysical Observatory of the  Russian Academy of Sciences (SAO RAS) and 1m Schmidt telescope of the  Byurakan Astrophysical Observatory (BAO) of the National Academy of Sciences of Armenia for deep spectral and imaging  data. Throughout this study, we adopted the Mrk 738 redshift $z=0.0673$ that gives in the standard $\Lambda$CDM cosmology (H$_0$ = 68 km s$^{-1}$  Mpc$^{-1}$, $\Omega_m=0.31$) the luminosity distance	318 Mpc and the    scale   1.35 kpc arcsec$^{-1}$  according the   NED database\footnote{\url{http://http://ned.ipac.caltech.edu/}}.

\begin{table}[h]
 \caption{Log of optical  observations at the 1m (BAO), 2.5m (MaNGaL) and 6m (SCORPIO-2) telescopes.
\label{tab:spec_data}}
		\begin{tabularx}{\textwidth}{rccccc}
		\toprule
	\hline
Data set        & Date     & $T_{exp}$, s      & $\beta$, $''$   & Sp. range      & $FWHM$, \AA     \\ 
\hline
\multicolumn{6}{c}{Direct imaging} \\    
  MaNGaL    & 2022 Apr 07 & 2000 &   1.8     &  \OIII$\lambda5007$   & 13 \\
  MaNGaL    & 2022 Apr 07 & 1200 &   1.8     &  Continuum    & 13 \\
  MaNGaL    & 2022 Apr 25 & 3600 &   1.6     &  \OIII$\lambda5007$   & 13 \\
  MaNGaL   & 2022 Apr 25 & 3200 &   1.6     &  Continuum    & 13 \\
    BAO   & 2023 May 16 & 2400 &   2.7     &  $r$-sdss    &  \\
\multicolumn{6}{c}{Long-slit spectroscopy}\\
 SCORPIO-2   $PA=34^\circ$ & 2023 Mar 14 & 2400 & 2.4   & 3650--7300 \AA  & 4.5   \\
 SCORPIO-2   $PA=95^\circ$ & 2023 Mar 14 & 2400 & 2.0   & 3650--7300 \AA  & 4.5   \\
 SCORPIO-2   $PA=137^\circ$ & 2023 Mar 15& 2400 & 1.8   & 3600--8550 \AA  & 7.0   \\
\hline
   			\bottomrule
		\end{tabularx}
\end{table}

\begin{figure*}
  \centering
 \includegraphics[width=\textwidth]{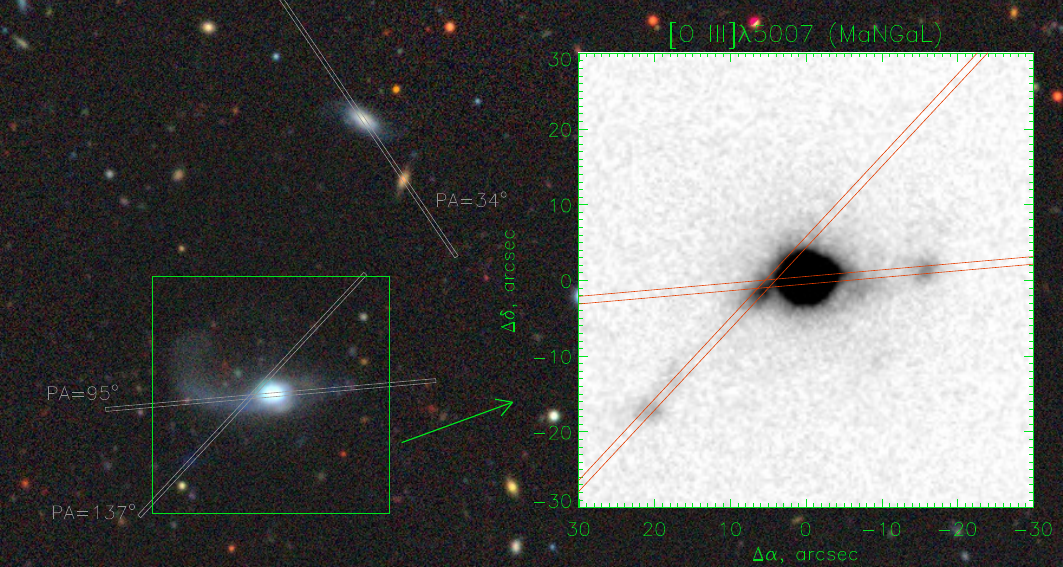}
\caption{DESI Legacy survey  image of Mrk~783 {(the  combination of images in $grz$ filters)} together with MaNGaL \OIII{} map. The SCORPIO-2 slit positions are shown by grey and red lines. 
}\label{fig1}
   \end{figure*}

\section{Observations and data analysis}

\label{sec:phot}

Images in the \OIII{} emission line were obtained in March 2022 at the 2.5m CMO SAI MSU telescope \citep{Shatsky2020}  with the tunable filter photometer  MaNGaL (Mapper of Narrow Galaxy Lines). 
This instrument is  using a scanning Fabry-Perot interferometer as a narrow-band  filter ($FWHM\approx13$\,\AA). The filter was subsequently centered  on the redshifted \OIII$\lambda5007$ emission line and on the continuum shifted in 80\,\AA{} from the line. The instrument description and data reduction steps are described in \citet{Mangal}, 
the log of observations is given in Tab.~\ref{tab:spec_data} where $T_{exp}$, $\beta$ and $FWHM$ are a total exposure, seeing value and spectral resolution correspondingly. The field of view (FoV) was $5.6'$ with the scale $0.33''/$px. 
The images obtained in different nights were aligned using the astrometric calibration via the astrometry.net software \citep{Lang2010AJ....139.1782L}.  The final emission line image  after   continuum subtraction is shown in Fig.~\ref{fig1}. We have mapped \OIII$\lambda5007$ emission  to the surface brightness  $\sim3.6\cdot10^{-17}\,\mbox{erg}\,\mbox{s}^{-1}\mbox{arcsec}^{-2}$ with   the signal-to-noise ratio $S/N\geq 3$.

\begin{figure*}
\centering
\includegraphics[width=\textwidth]{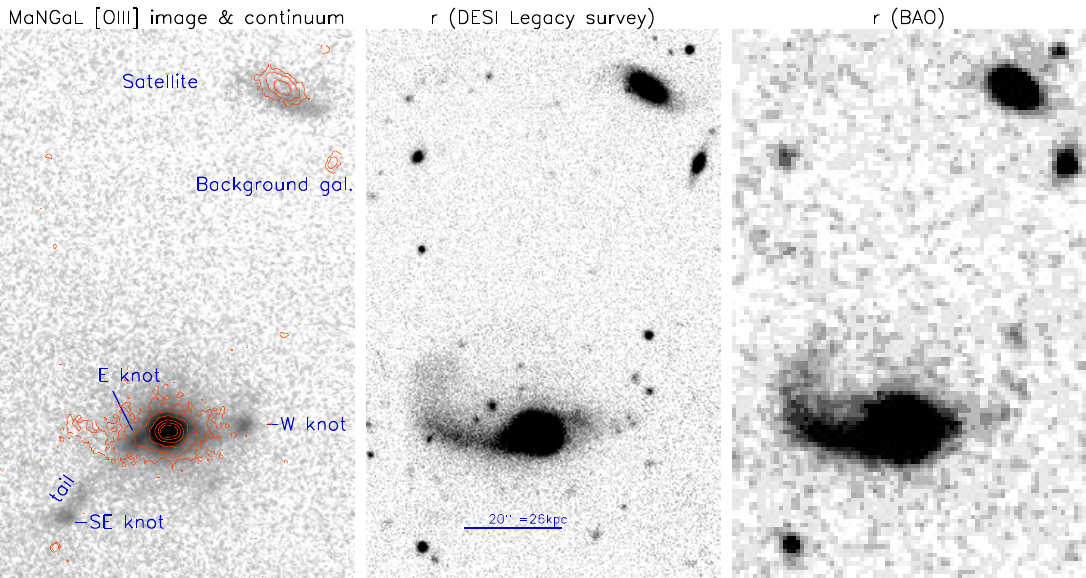}
\caption{Mrk 783 with the satellite SDSS J130257.20+162537.1. From left to right: the  \OIII{} emission line map with contours of MaNGaL image in  the continuum, the main emission knots are labeled; the DESI Legacy survey $r$-band image; deep $r$-band image from 1m BAO telescope.}
\label{fig:deep}
\end{figure*}

The \OIII{} image reveals an extended  system of emission knots and diffuse ionized gas surrounding the main galaxy. The brightest external features are (Fig.~\ref{fig:deep}): (i) `SE knot' at the projected distances $r=26-29''$ from the nucleus that is a part of emission `tail'; (ii) `E knot' -- the bright region at $r=5-9''$ to the east that is root of the `tail' and (iii) `W knot' at $r=13-18''$ . Some of regions listed  above   are visible in DESI Legacy survey \citep{Legacy2019AJ....157..168D} as a faint blue structures (Fig.~\ref{fig1}),   
SE knot was already found in the Magellan spectra \citep{Congiu2017b}.  Also a significant  \OIII{} emission was detected in the galaxy SDSS J130257.20+162537.1 located at $\sim73''$ (99 kpc) projected distances to the north of Mrk~783 (hereafter SDSS~J1302+1625 or `the satellite'). 

In order to study in details faint stellar structures outskirts of Mrk 783 we performed deep imaging in the $r$-SDSS filter with the 1m Schmidt telescope of the BAO\footnote{The galaxy Mrk~783 was discovered with this telescope in 1976 \citep{Markarian1976}}. The updated $4K\times4K$ Apogee (USA) liquid-cooled CCD
camera was used as a detector with the pixel scale $0.86''$ and FoV of $\sim1^\circ$.  
The detailed description of the telescope, the photometer and data reduction steps are given in \citet{Dodonov2017AstBu..72..473D}. 
The $r$-band image was calibrated to the  magnitudes by using the DESI  Legacy Surveys DR10 on-line  photometric catalogue\footnote{\url{https://datalab.noirlab.edu/}} of sources in the observed field. At the signal to noise value $S/N=3$ we reached the surface brightness limit $25.3~mag\,arcsec^{-2}$ that is on $\sim0.6$ mag deeper than $r$-band  DESI Legasy survey image (Fig.~\ref{fig:deep}).

The spectral observations were carried out at the prime focus of the SAO RAS  6m telescope in the long-slit mode of  the SCORPIO-2  focal reducer \citep{AfanasievMoiseev2011} providing the spatial scale $0.39''/$px along the  $1''\times6.4'$ slit. 
Other parameters are listed in the Tab.~\ref{tab:spec_data}. We put the slit along the most interesting features appearing in \OIII{} map: the slit with position angle $PA=34^\circ$ crossed the possible companion  galaxy SDSS~J1302+1625   and fainter red galaxy SDSS J130256.50+162521.8, the slit with $PA=95^\circ$ crossed the Mrk~783 nucleus together with W and E knots, whereas the slit $PA=137^\circ$ exposed the tail with E and SE knots (see.~\ref{fig1}).  
{The spectrophotometric calibration was based on the observations of the  standard star  BD+75~325. Also the  spectrum of this O-type star was used to correction of the galaxy spectra on the telluric absorption: O$_2$ $B$-band (686--698 nm) and H$_2$O band at 710--730 nm. The last one   affects the \SII$\lambda\lambda6717,6731$ redshifted lines of Mrk~783.}

\begin{figure*}
  \centering
\includegraphics[width=\textwidth]{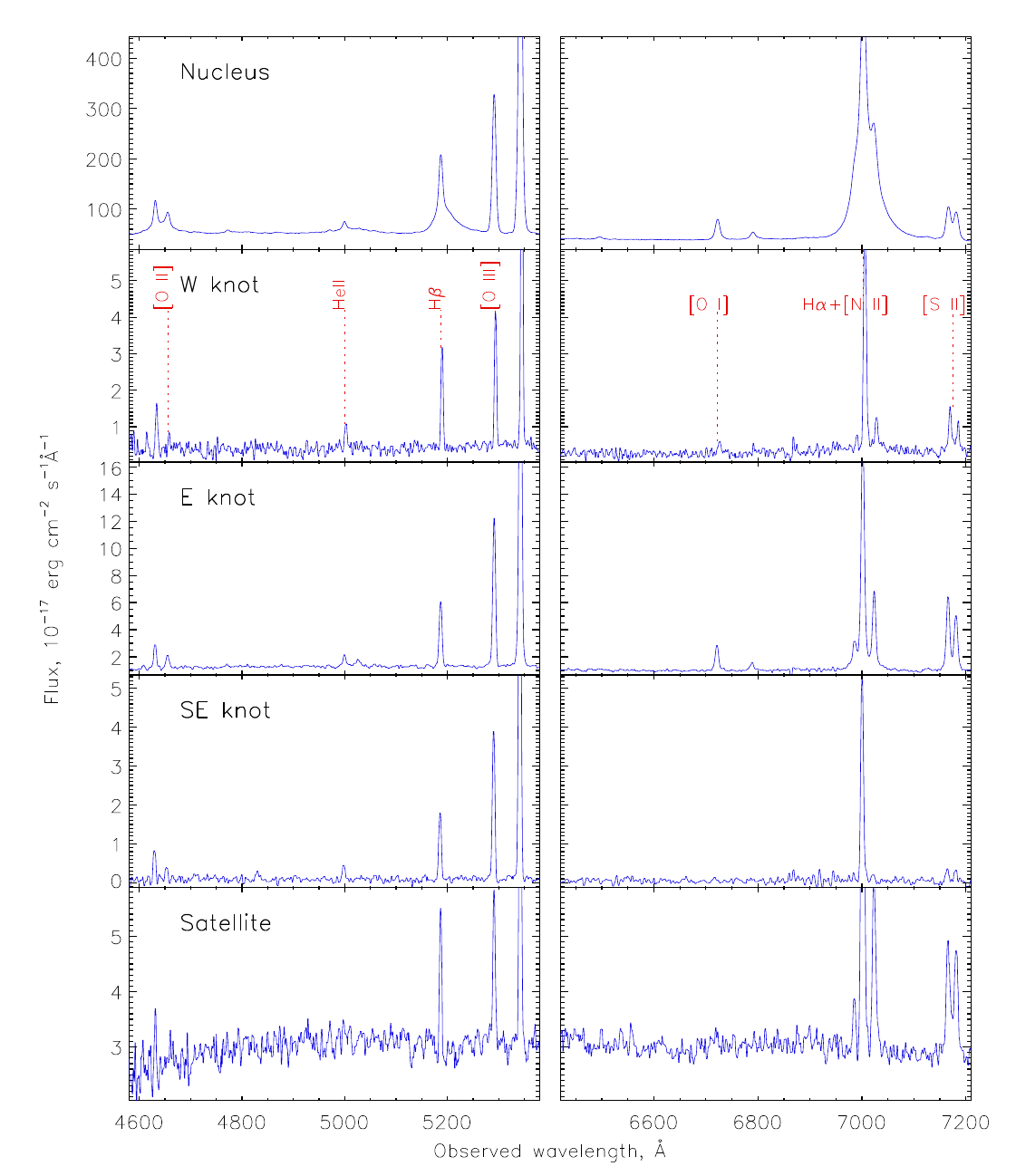}
\caption{SCORPIO-2 spectra of the Mrk 783 nucleus, W, E and SE emission knots  and the satellite. All spectra are integrated in $6''$ aperture. The main emission lines are labeled.  }\label{fig:spec}
   \end{figure*}

\begin{figure*}
  \centering
\includegraphics[width=\textwidth]{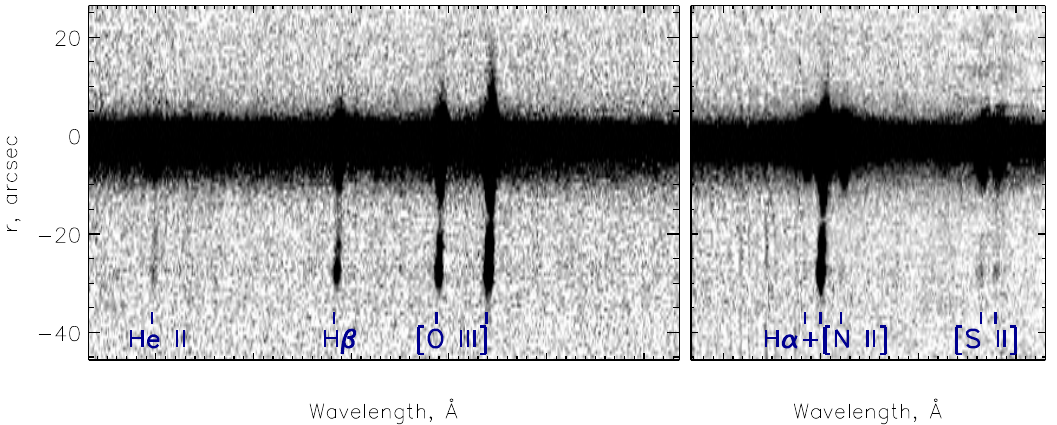}
\caption{Fragments of the 2D spectrum along $PA=137^\circ$
in the blue and red ranges  binned to a scale of 0.8$''$/px. }\label{fig:spec2}
   \end{figure*}

\section{Ionized gas properties}
\label{sec:gas}

The integrated spectra of the emission knots mentioned above together with  nuclei  of the both galaxies are shown in Fig~\ref{fig:spec}. The nuclear spectrum demonstrates the same properties detected in previous studies, including SDSS archival data\footnote{\url{https://dr18.sdss.org/optical/spectrum/view?plateid=2603&mjd=54479&fiberid=259}}: a broad blue-shifted component of the Balmer emission lines and Sy-like flux ratio without bright iron emission. The both  low-  and high-excitation emission lines in external  knots are narrow with a  single-component structure. Also we detected here high-excited \HeII$\lambda4686$ emission with its  relative intensity  similar to the nuclear source: \HeII/\Hb$\approx0.25$. Fig.~\ref{fig:spec2} clearly demonstrates the extended emission in \HeII{} along the SE gaseous tail up to the projected distance $r\approx30''$ (41 kpc). The spectrum of the satellite (Fig.~\ref{fig:spec}, bottom panel) corresponds to intermediate objects between AGN and starburst (see Sec.~\ref{sec:sat}) with narrow Balmer, \SII,\NII{} and \OIII{} lines  without helium emission.

\begin{figure*}
\centerline{
\includegraphics[width=0.33\textwidth]{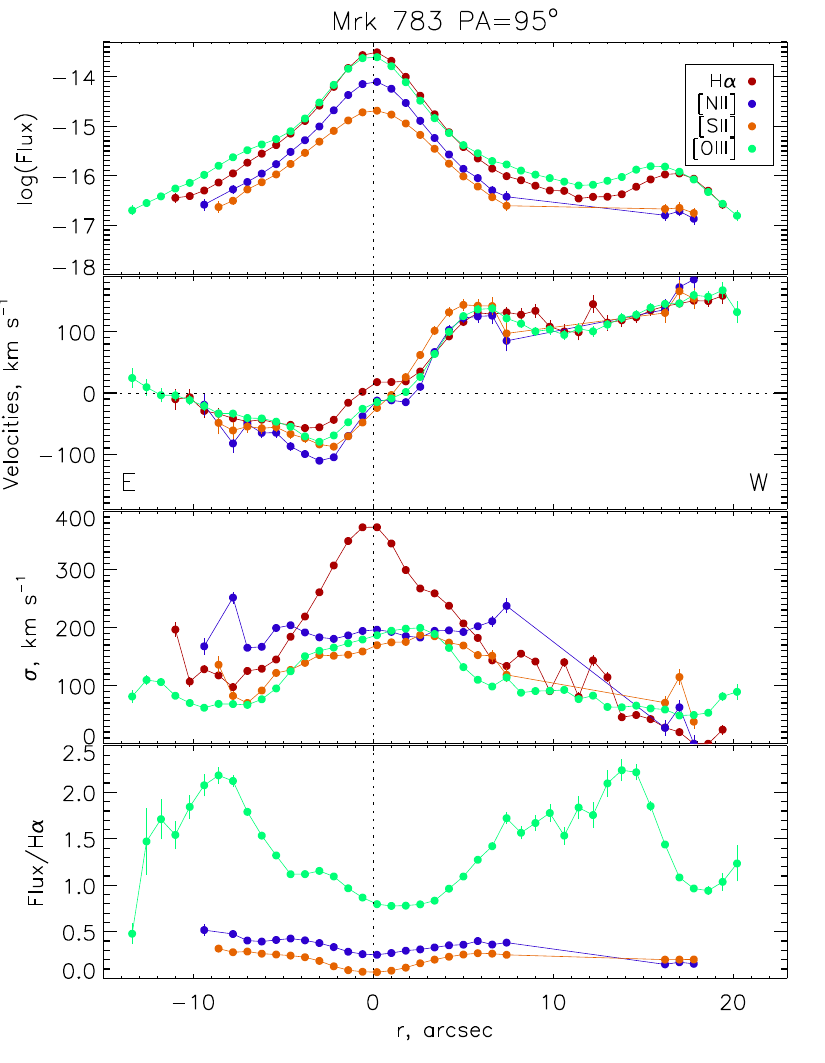}
\includegraphics[width=0.33\textwidth]{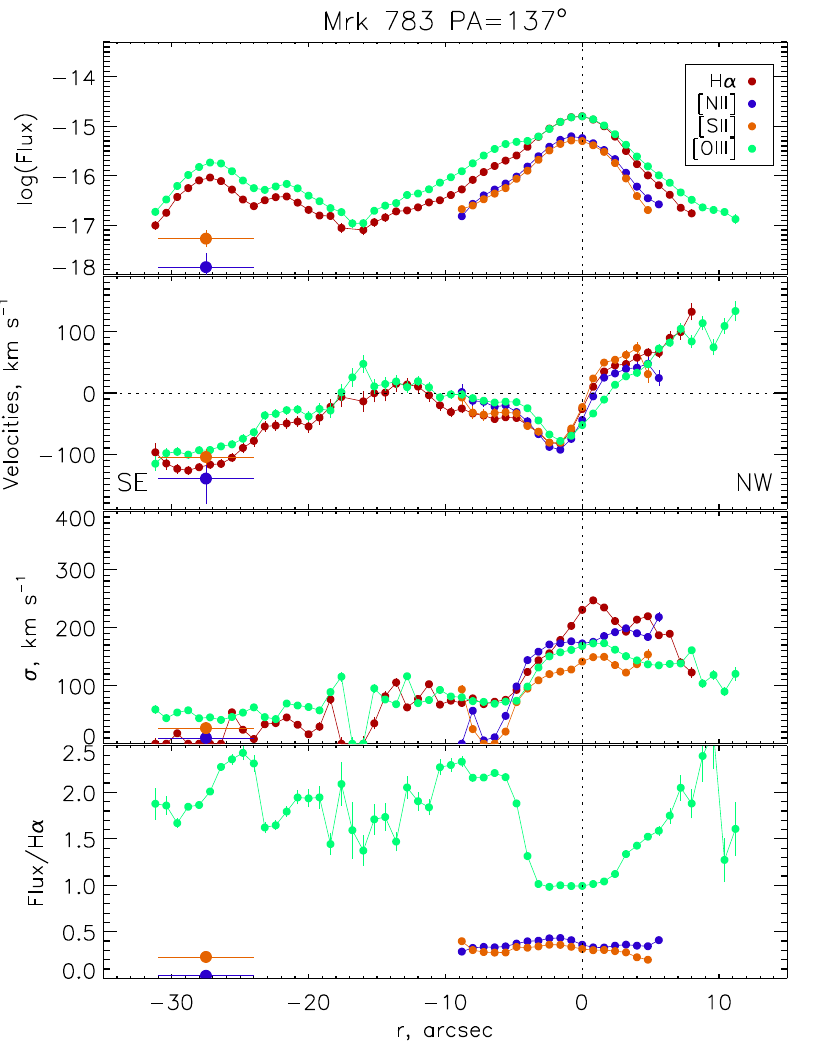}
\includegraphics[width=0.33\textwidth]{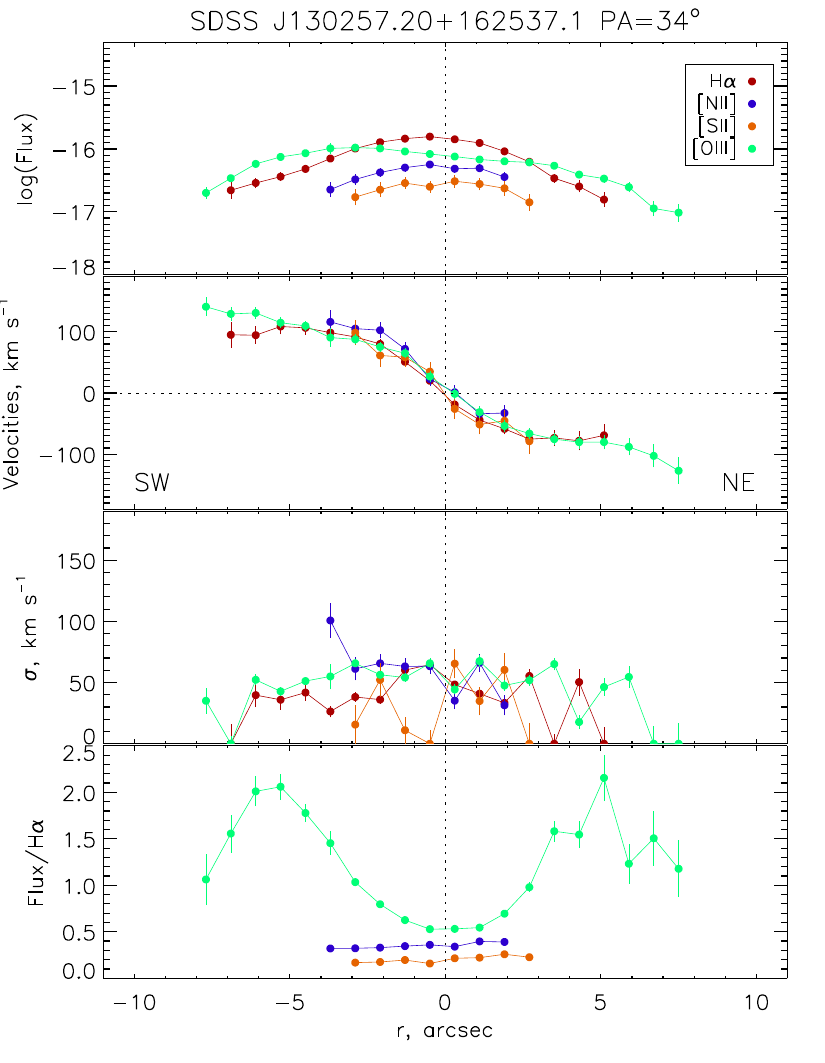}
}
\caption{The changes of the main emission lines parameters along $PA=95^\circ$ (left) and $PA=137^\circ$ (center) and $PA=34^\circ$ (right). From top to bottom : the surface brightness, line-of-sight velocities (the systemic velocity $20100\kms$ and $20038\kms$ were subtracted for the Mrk~738 and  SDSS~J1302+1625  correspondingly), velocity dispersion corrected on the instrumental width and flux ratio relative to the \Ha. The largest symbols in the central panel correspond to the integrated values in the SE knot. The   point $r=0$ corresponds to the maximal  emission in the stellar continuum.}
\label{fig:lines}
\end{figure*}

The parameters of the  emission lines after subtraction of continuum interpolated  by cubic spline  (integrated flux, line-of-sight velocity and velocity dispersion corrected on the instrumental broadening) were estimated  using a single-Gaussian fitting and  shown in the Fig.~\ref{fig:lines}.  

In the inner few kpc ($|r| \le 4''$) our long-slit data reveal  features probably related with AGN outflow influence on the surrounding gas: a high  velocity dispersion in the  \Ha{} line (including a spread of light from the broad line region),  a significant difference between velocities in the forbidden and Balmer lines at ($r\approx-3''$ in $PA=95^\circ$, {see Fig.~\ref{fig:lines}}), a peak of negative velocities near the E-knot ($r\approx-2''$ in $PA=137^\circ$).  In contrast   with a circumnuclear region, the external gas in and around Mrk 783  is  dynamically cold: the observed velocities in the forbidden and Balmer lines are in a good agreement within the errors; the typical velocity dispersion is   $\sigma<100-150\kms$. The line-of-sight velocity curve   along  $PA=95^\circ$ at $r>3''$ seems like a typical flat rotation curve of a disk galaxy (the slit lies near its photometric major axis). In this case the  falling velocities in the opposite (eastern) side of this curve should be correspond to rotation in the stellar tidal tail clearly visible in broad-band images which can be partly off-plane. The amplitude of observed velocities along $PA=137^\circ$ is about $100\kms$ that is in an agreement with gas rotation on orbits slightly  inclined to the main galaxy disk.

\begin{figure*}
  \centering
\includegraphics[width=\textwidth]{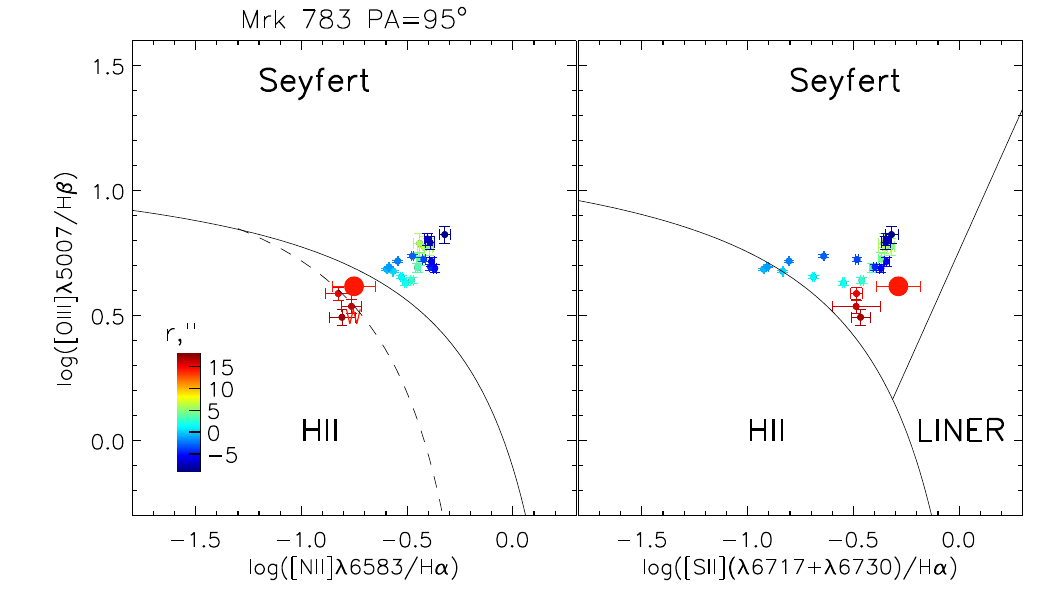}
\includegraphics[width=\textwidth]{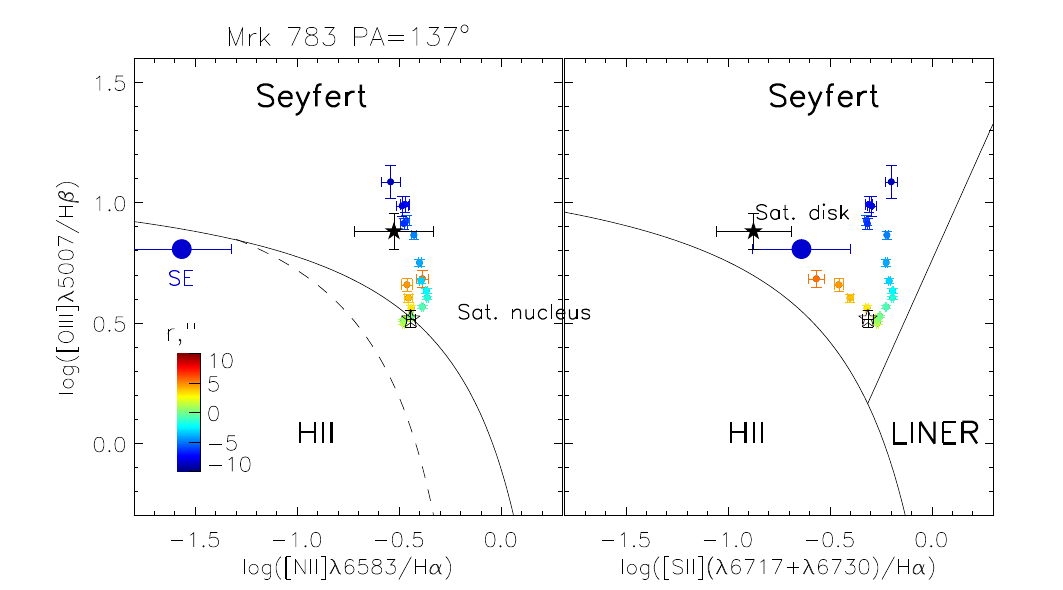}
\caption{Line ratio (BPT) diagrams  for the binned spectra along $PA=95^\circ$ (top) and  $PA=137^\circ$ (bottom). Dividing lines between HII regions, AGN and composite nucleus are taken from \citet{Kewley2001ApJ...556..121K, Kauffmann2003MNRAS.346.1055K}. 
{Different colours correspond to radial distances along the slit according to the scale box.} 
Large red and blue points correspond to the integrated values for the W and SE knots. The open and filled black stars in the bottom panels show line ratios for the nuclear and external regions of the satellite.}\label{fig:BPT}
\end{figure*}

The  \OIII{} to \Ha{}  flux ratio along $PA=137^\circ$ has a more or less constant  value  \OIII$\lambda5007$/\Ha$\approx2$ in a very large distant range $r=-30...-5''$ implying  the same source of the gas excitation along the SE tail.  Fig.~\ref{fig:BPT} shows the diagnostic emission-line flux ratio diagrams  
\citep[BPT, after Baldwin et al. ][]{Baldwin1981PASP...93....5B} for the  Mrk~783 EELR. In contrast with the previous spectroscopic observations \cite{Congiu2017b} where only upper limits for the  \NII/\Ha{} ratio were presented   for the most distant SE-knot, we able to detect in this region both relatively faint \NII{} and \SII{} doublets of emission lines (Fig.~\ref{fig:spec2}).

On the diagram \OIIIHb{} vs. \SIIHa{} (Fig.~\ref{fig:BPT}, right) all points occupied an area corresponded  to  the AGN-type ionization, whereas in the case of  \OIIIHb{} vs. \NIIHa{} plot (Fig.~\ref{fig:BPT}, left) the most distant regions locate near the border between the  \HII{} and composite excitation (W knot) or even in the \HII{} area (SE knot). On the other hand, the high value of   \HeII/\Hb{} lines ratio in  these regions suggest (as well as in a bulk of EELR) the  photoionization by hard UV  continuum from AGN than by young OB stars. In this case, a decreasing of \NIIHa{} ratio for the  external regions could be caused  by a relatively low gas chemical  abundance. Indeed, numerical calculations by \citet{Bennert2006A&A...446..919B} for the typical NLR exhibit  the similar behaviour of  points on the BPT-diagramms, if we accept a metal abundance values  $Z=0.1-0.3 Z_\odot$ for the W-knot and $Z\approx0.05$ for the NW-knot.

Therefore the observed emission line ratios (\HeII/\Hb{} and BPT diagrams) suggest that AGN radiation  is dominated source of the gas ionization up to projected distance 41~kpc from the nucleus.

\section{The galaxy environment and the satellite}
\label{sec:sat}

As we have already noted in Sec.~\ref{sec:intro}, Mrk~783 was considered as an interacting  galaxy having tidal tail and a possible secondary nucleus.  The SCORPIO-2 spectrum along $PA=34^\circ$ reveals  that the brightest and nearest candidate to a possible companion --- the galaxy   SDSS~J1302+1625 (`the satellite') has a systemic velocity $V_{sys}=20038\pm15\kms$ that  deviates only on $72\kms$ from the velocity of Mrk 783 nucleus according our estimations along $PA=34^\circ$ ($20100\pm20\kms$, it is a mean value in the \Ha,\NII and \OIII{} lines) or on $165\kms$ if we accepted the  velocity of  Mrk 783 nucleus according SDSS data in NED. These small velocity difference implies that  Mrk~783 and SDSS~J130257.20+162537.1 is  a  gravitationally bound pair. The slit $PA=34^\circ$ also crossed a smaller galaxy in 20'' from the satellite --- J130256.50+162521.8, its spectrum corresponds to a distant galaxy    redshifted at  $z\approx0.21$. It is marked in Fig.~\ref{fig:deep}  as a `background galaxy'.
We also found no other Mrk~783 companions by looking at the list of spectral and photometric redshifts for all NED objects up to projected distance 500 kpc ($6.2'$).

Can the  satellite SDSS~J1302+1625 create the observed peculiar morphology of Mrk 783?  
The most prominent tidal tail  expands in E and NE directions up to projected distance $26''$ (35 kpc) from the Mrk 783 center. The both DESI Legacy survey and deep BAO images (Fig.~\ref{fig:deep}) demonstrate an absence of any faint tidal structures between the main galaxy and the satellite at least up to surface brightness $25.3~mag\,arcsec^{-2}$ in the $r$-band (Sec.~\ref{sec:phot}). The comparison of  SDSS DR18 red  magnitudes of  Mrk~783 ($m_i=15.34$, $m_z=15.67$) with those for the satellite ($m_i=17.67$, $m_z=17.46$) gives the ratio  in their luminosity and hence stellar mass $1/5$--$1/9$. This  low  ratio corresponds  to the case of minor merging without significant perturbation of the main galaxy. Moreover  in contrast with the main galaxy, the satellite seems unperturbed in both   morphology and  internal kinematics: it has a symmetrical rotation curve of the ionized gas (Fig.~\ref{fig:lines}). All the facts listed above together with a possible sign of the secondary nucleus \citep{Congiu2020} suggest that the observed peculiar morphology of Mrk 783 was caused by previous external event (merging with a  companion) rather than with the low-massive satellite in $\sim100$ kpc away.

The ionized gas properties of  the satellite are  intriguing. As we already mentioned in Sec.~\ref{sec:gas} its nuclear spectrum corresponds to  a starburst galaxy.  The emission line ratios correspond to the border between AGN and HII regions in BPT diagrams (open black star in Fig.~\ref{fig:BPT}, bottom panels). However the excitation  of \OIII{} emission increase dramatically to the galaxy outskirts ---  the ratio  \OIII/\Ha{} reaches $\approx2$ that is similar with the Mrk 783 EELR (Fig.~\ref{fig:lines}, bottom). The corresponded points on the BPT-diagramms moves upward in the AGN area, {in the region occupied by Mrk~785 EELR }
 (the filled black star in Fig.~\ref{fig:BPT}, bottom).  This fact implies   that the outer part of the satellite disk is also ionized by the Mrk~783 AGN. 

The second argument in favor of the external origin of the gas ionization in the satellite disk comes from the \OIII{} emission distribution according MaNGaL data (Fig.~\ref{fig:deep}). It is clearly seen that highly ionized gas in the satellite placed asymmetrically relative its nucleus  --- we observed  it mostly on the side nearest to  the Mrk 783.

{We tried to estimate the electron density $n_e$ from the density-sensitive \SII{} doublet flux ratio  $R=F($\SII$\lambda6717)/F($\SII$\lambda6731$) using the diagnostic  equations from \citep{Proxauf2014A&A...561A..10P} for $T_e=10^4$~K. The values of   $R$ derived from the integrated spectra of W-knot, SE-knot and of the outer part  of  SDSS~J1302+1625 (in the range $r=3.5$--$9''$) are $1.41\pm0.06$, $1.03\pm0.09$ and $0.86\pm0.19$  that corresponds to $n_e=20\pm50$, $400\pm140$ and $770\pm500\,\mbox{cm}^{-3}$ with the $1\sigma$ level. Preliminary we can conclude that the ionized gas density is significantly higher in  the SDSS~J1302+1625 disk than in Mrk~783 EELR,  but the uncertainty in $n_e$ estimation is too great for  more detailed analysis.  New deeper spatially-resolved  spectroscopic data   are needed to better understand the ionization properties of the  SDSS~J1302+1625 outskirts. }

\section{The energetic budget}
\label{sec:energy}
External  EELRs are considered as a good probe to  study the history of AGN radiative output on the time scale $10^4$ -- $10^5$ yr (it corresponds to light-travel time to the gaseous clouds). The well-known prototype is Hanny’s
Voorwerp, a  cloud of highly ionized gas near  the spiral galaxy IC 2497. The detailed comparison of the ionized gas properties and AGN luminosity clearly demonstrates that the radiation associated with a nuclear activity significantly fall at least two orders in the last   $\approx10^5$ yr  \citep{Lintott2009MNRAS.399..129L, Schawinski2010ApJ...724L..30S,Keel2012AJ....144...66K}. New examples of `fading' AGNs were discovered  in follow-up spectroscopic observations of EELR  candidates found  in  SDSS broad-band images by  volunteers of the   Galaxy Zoo citizen-science project \citep{Keel2012MNRAS.420..878K,Keel2017ApJ...835..256K}, as well as in the data collected in surveys based on  narrow-band \OIII{} imaging \citep{Keel2022MNRAS.510.4608K} or  integral-field spectroscopy  \citep{French2023ApJ...950..153F}. The ionized gas clouds around Mrk~783 including the disk of the companion galaxy allows as to use the same technique for  estimation of its ionization budget.

We calculated the ionizing luminosity ($L_{ion}$) required to creation the distant emission knots  with the current bolometric luminosity of AGN  ($L_{AGN}$) using the approach proposed in the paper cited above and briefly described below.
The upper limit of the AGN flux absorbed by dust was estimated  as a sum  of fluxes in   the far infrared (FIR, the wavelength range 42-122$\,\mu m$) according the  {\it Infrared Astronomical Satellite} (IRAS) point-source catalogue data \citep{IRAS} and mid-infrared (MIR, the wavelength range 3.4-42$\,\mu m$) data from the {\it Wide-filed Infrared Survey Expoler} \citep[WISE,][]{WISE}. We evaluated the {\it IRAS} luminosity ($L_{FIR}$) using a standard linear combinations of the flux in the 60 and 100 $\mu m$ bands in the similar way as described in \citet{Keel2012MNRAS.420..878K}. For the {\it WISE}  luminosity ($L_{MIR}$) we use a power-low approximation of the flux values obtained  from the NED  in the four bands 3.4, 4.6, 11.6 and 22.08$\mu m$  in the same manner as in \citet{Keel2017ApJ...835..256K}. For the  unobscured AGN luminosity we use an empirical equation from \cite{Keel2022MNRAS.510.4608K}: $L_{unobs}=340\,L_{OIII}$, where $L_{OIII}$ is a nuclear luminosity in the \OIII$\lambda5007$ emission line based on the nuclear flux derived from 2D Moffat fitting of the MaNGaL image.   
In this case:
\begin{equation}
L_{AGN}=L_{FIR}+L_{MIR}+L_{unobs}
\label{equ:agn}
\end{equation}

The lower limit of the nuclear source luminosity required for the ionization of the gas region having the \Hb{} flux $F$(\Hb) occupying an angle $\alpha$  was estimated according the equation proposed in \citep{Keel2012MNRAS.420..878K}:
\begin{equation}
\begin{array}{l}
L_{ion}=1.3\cdot10^{64}z^2F(\mbox{H}\beta)/\alpha^2\\
\alpha=2\arctan(W/r)\\
\end{array}
\label{equ:ion}
\end{equation}
where $z$ is a galaxy redshift  and $\alpha$ is a projection of a solid angle under which the considered region is viewed from the nucleus.  Here $r$ -- is a projected distance from the nucleus, whereas $W$ is a radius of this region.

The  quantities related to the energy balance based on the eq. (\ref{equ:agn})--(\ref{equ:ion}) are listed in the Table~\ref{tab:balance}. For the SE-knot we integrated the \Hb{} flux from the SCORPIO-2 spectra along $PA=137^\circ$ in the range $r=24-31''$,  $W=0.5''$ that is a half of the slit width. The eq. (\ref{equ:agn}) is written for the slit oriented radially to the AGN, that is true for $PA=137^\circ$. But it is not valid for the slit orientation $PA= 34^\circ$, passing through the satellite. To account for this, we  used  the same technique as described in \citet{Keel2022MNRAS.510.4608K} for the similar observations of the external gas cloud near NGC~5514: i.e. we multiply the square of the emission edge of the satellite disk according   \OIII{} MaNGaL map (Fig.~\ref{fig:deep}, left) on the mean surface brightness of the external part of the disk at $4-9''$ from the satellite nucleus according   spectra along $PA=34^\circ$. In this case,  $W=7''$ (the radius of the emission region) and $r=72''$ (the distance from the AGN).

The ratio $a_{ion}=L_{ion}/L_{AGN}$ could be considered as an indicator of a long-term fading of AGN radiation or a difference of  dust obscuration of this radiation between the direction to an observer and to EELR. Usually a fading AGN has $a_{ion}\ge3$ \citep{Keel2012MNRAS.420..878K,  Keel2022MNRAS.510.4608K}. However in the Mrk 783 this value is significantly smaller: $a_{ion}\approx1$ in the SE-knot and even 0.12 in the outer disk of the companion galaxy. Of course, the real value $a_{ion}$  could be higher, because the eq. (\ref{equ:agn}) gives an upper bound of the active nucleus infrared luminosity and includes a fraction related with a dust  heated by a star formation in the galaxy. Whereas, the eq.~(\ref{equ:ion}) gives a low bound of the ionizing flux absorbed by the clouds, because it depends on spatial resolution and optical thickness in the Lyman continuum \citep[see the discussion in ][]{Keel2012AJ....144...66K}. A real spatial geometry of the system is   unknown. Nevertheless we have no arguments in favour of  significant fall  AGN radiation in last 0.1--0.3 Myr, that corresponds to a projection of travel-light time from the Mrk~783 nucleus to the SE-knot and to the disk of SDSS~J1302+1625.

\begin{table}[h]
 \caption{The energy balance between observed AGN output and requared photoionization to power the emission clouds
\label{tab:balance}}
		\begin{tabularx}{\textwidth}{rlll}
		\toprule
	\hline
Quantities           & AGN      &  SE-knot   & Satellite \\ 
\hline
FIR  luminosity  $L_{FIR},\ergs$   &  $2.09\cdot10^{44}$     & & \\
MIR luminosity $L_{MIR},\ergs$ &  $1.89\cdot10^{44}$      & & \\
Unobscured ionizing luminosity $L_{unobs}\ergs$  &  $5.51\cdot10^{44}$      & & \\
Total ionizing luminosity $L_{AGN},\ergs$   &   $9.49\cdot10^{44}$   &  &  \\
\hline
 Viewing angle  $\alpha$, $^\circ$   &       & 2.4 & 11.1 \\
Cloud \Hb{} flux, $\ergs\,\mbox{cm}^ {-2}$      &       & $1.14\cdot10^{-16}$ &  $2.4\cdot10^{-16}$  \\
$L_{ion}\ergs$       &    &  $1.18\cdot10^{45}$  & $1.16\cdot10^{44}$  \\
$a_{ion}=L_{ion}/L_{AGN}$ & & 1.24 & 0.12   \\
      \bottomrule
		\end{tabularx}
\end{table}

\begin{figure}
\centering
\includegraphics[width=0.5\textwidth]{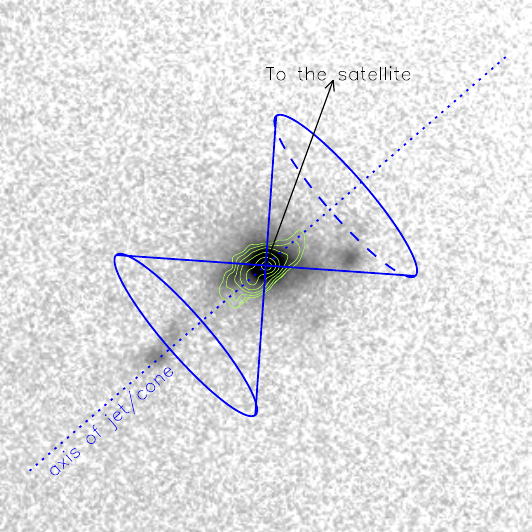}
\caption{The \OIII{} emission line image with contours of radio continuum at 5 GHz according \citet{Congiu2017a}. Proposed orientation of ionization cone is shown.}
\label{fig:layout}
\end{figure}

\section{Discussion }

It is generally accepted that  large-scale EELRs around AGN can be  formed in  two main ways: via nuclear outflow driven by the  kinetic power of  radio jet and/or superwind in radio-loud AGN \citep{Harrison2014MNRAS.441.3306H,Rupke2011ApJ...729L..27R} or as a result of ionization of the pre-existing gas surrounding  radio-quiet Seyfert galaxies \citep{Keel2012MNRAS.420..878K,Keel2022MNRAS.510.4608K}. The combination of the both cases is also possible, including a relic structures from the  previous activity  episodes \citep{Morganti2017NatAs...1..596M}.
The radio to optical luminosity ratio puts the  Mrk~783   between radio-quiet and radio-loud AGN, the galaxy contains both an inner jet and extended diffuse radio structure \citep{Congiu2017a}. What is an origin of this EELR? 

Our spectroscopic data  clearly manifest  that the possible sign of jet outflow in the ionized gas kinematics appears only in the circumnuclear region  ($r<5$ kpc, Sec.~\ref{sec:gas}). Whereas the more external gas exhibits properties similar with  the external off-plane gas and tidal debris ionized by AGN in other galaxies \citep[][and references therin]{Keel2012AJ....144...66K, Keel2022MNRAS.510.4608K}: 
\begin{itemize}
    \item The  flux ratio of the most indicative emission lines (BPT-diagrams, high \HeII/\Hb{})   corresponds  to the ionization by UV-continuum of AGN rather than by shocks related with jet or outflow.
    \item The quiet kinematics of gas clouds (rotation on circular orbits, relatively low velocity dispersion)  {also indicates  tidal induced motions or an external gas accretion}. 
\end{itemize}

In their  paper based on the Magellan spectroscopy \citet{Congiu2017b} assumed that the ionized gas excitation in the Mrk~783 EELR is related with the extended radio structure. However the comparison of their synchrotron radio isocontours with   MaNGaL \OIII{} image manifests that only SE and E  knots  are more or less aligned with a direction of central radio structure, wheres the W-knot and the surrounding gas  are not coincide with the radio contours (Fig.~\ref{fig:layout}). On the other hand, the gas excitation and kinematics  are similar in all considered  structures that implies a common source of its ionization.  On   Fig.~\ref{fig:layout} we draw  a location of possible bi-symmetric ionization cones with axis aligned $PA=131^\circ$ according to  a large-scale jet orientation \citep{Congiu2017a}. The most of EELR structure lies inside the cones if  we accepted the projected value of cone's open angle $90^\circ$. This value is in agreement with the mean value of $70^\circ$ for other ionization cones \citep{Keel2019MNRAS.483.4847K} if the projection effect will be taken into account. 

The proposed orientation of the AGN cones  allows to explain the ionization of the external parts of gaseous disk  in the  satellite galaxy (see the arrow in Fig.~\ref{fig:layout}).  As minimum a half of its disk that is close to the Mrk 783 is also ionized by the AGN. The effect of cross-ionization by  companion's  AGN have been already found in several  galactic pairs, however the characteristic separation was about 15-20 kpc \citep{Moran1992AJ....104..990M,Merluzzi2018ApJ...852..113M,Keel2019MNRAS.483.4847K}. In the case of Mrk~783 system the projected distance between companion's nuclei is about 100 kpc that is perhaps the largest known today.

In the sec.~\ref{sec:sat} we presented arguments that the disturbed morphology of Mrk~783 and its tidal structures is not related with a distant companion galaxy. Most likely we  observe the result of merging with a gas-rich dwarf galaxy. In this case the most of external gas in EELR came from a tidal destroyed     low-metallicity companion ($0.05-0.1Z_{\odot}$ according the low \NIIHa{} ratio, sec.~\ref{sec:gas}). The stellar tidal structure is aligned with gaseous one on the west from the galaxy, but NE emission tail has no stellar counterpart. This stellar-gaseous spatial misalignment is  also observed in some AGN interacting galaxies in  which  EELRs are not spatial coincide with  stellar tidal structures (for instance: NGC 5278/9 \citep{Keel2019MNRAS.483.4847K} or a spectacular local example ---  NGC 5194/95 \citep{Watkins2018ApJ...858L..16W}). It is not surprising because in \OIII{} we have detected only high-ionized fraction of a whole gaseous structure, whereas a distribution of cold \HI{} can be revealed only by radio observations.

Interesting to note that many properties of Mrk~783  are similar to those observed in  IC 2497 with Hanny’s Voorwerp nebula (Sec.~\ref{sec:energy}): both are post-interacting disk galaxies having detached AGN-ionized EELR  at the spatial scale of tens kpc and a relic structure traces the previous phase of  radio jet activity. It  means that we possible caught both galaxies after the switching between different types of activity \citep{Fabian2012ARA&A..50..455F, Morganti2017NatAs...1..596M}:  radio-loud (i.e kinetic)  mode and   radio-quiet (i.e. radiation) mode.
  However in  IC 2497 the radio outburst occurred $~100$ Myr ago \citep{Smith2022MNRAS.514.3879S}, whereas a significant shortfall of
AGN ionizing radiation traced in Hanny’s Voorwerp nebula is dated as $\sim0.1$--0.2 Myr ago \citep{Lintott2009MNRAS.399..129L,Keel2012AJ....144...66K}. In Mrk~783 the age of the relic radio structure was not evaluated, whereas our calculations  of  the energetic budget (Sec.~\ref{sec:energy}) presented no evidences for  a significant decreasing of AGN radiation in last 0.1--0.3 Myr. Comparing these time scales we can speculate that the  Mrk~783 AGN ionizing radiation  will  be turned off  in the nearest future. Other words  it can be considered as   \textit{  `Hanny’s Voorwerp precursor'}. {However, a more in-depth study of the ionization balance in the disk of Mrk783 companion  is needed to prove this conclusion, because in the present work we   operated only  with integrated spectrum of the SDSS J1302+16 outskirts having a low surface brightness. Now we have performed only the first attempt to estimation of $L_{ion}$ in the EELR and the satellite disk, whereas the spatial resolution is crucial for using of eq.(\ref{equ:ion}), see \citet{Keel2012MNRAS.420..878K, Keel2017ApJ...835..256K}. Moreover,}
\citet{Congiu2020} found in Mrk~783 an inner  pc-scale   radio jet that is significantly misalignment with a kpc-scale structure. For the explanation they proposed two  scenarios: a jet precession, or    reactivation  after a period of inactivity. In this case the composition of recent activity episodes in Mrk~783 may be even more complex.  

\section{Conclusion}

Based on new  optical spectral and imaging observations we have studied the  distribution, kinematics and excitation of the ionized gas in the giant EELR of Mrk~783 galaxy as well as the properties of its environment. The deepest to date spectra of this area allows us to consider the gas ionization conditions up to 41 kpc from the galactic nucleus. Moreover, its ionization trace was found in the disk of the satellite galaxy. The main results are following:

\begin{itemize}
\item Mrk 783 forms a gravitationally bound pair with SDSS~J130257.20+162537.1 (the projected distance between their nucleus is $\sim99$ kpc). However the disturbed morphology and tidal structures are most likely caused by  merging 
with other pre-existing companion --- a gas-rich dwarf galaxy. 

\item Most of the gaseous structures detected in the emission lines  are ionized by the AGN radiation, but not by the radio jet.  

\item Part of the gas illuminated by the cone  belongs to the stellar tidal structure, but the most distant SE-knot is a part of external gaseous structure without stellar counterpart.  Gas in this region has a low metallicity ($0.05-0.1Z_{\odot}$ according the low \NIIHa{} ratio).

\item External regions  of the   satellite gaseous disk at the nearest side to Mrk 783  falls into the ionizing cone from the main galaxy  active nucleus. This fact makes the Mrk~783  system  perhaps the most extreme  example among nearby AGN galaxies of the cross-ionization  a galactic disk by a companion.

\item A comparison of  the ionizing luminosity  required  to create the most distant emission knots (including the satellite's disk) with the current bolometric luminosity of the nucleus indicates that there is  no  significant decreasing ionizing radiation during last 0.1--0.3 Myr.

\item  Mrk~783 can be considered   as a    `Hanny’s Voorwerp precursor', i.e. a galaxy that demonstrates  signs of sequential  switching from kinematics (radio-dominated) to radiation (ionization-dominated) AGN modes, in the moment before   falling of its ionization luminosity.
\end{itemize}

We hope that new multi-wavelength observations, first of all \HI{}  mapping in a radio domain, {and deep optical integral-field mapping of the both galaxies in the pair} allow us to better understand the spatial structure and evolution  of this galactic system.

\vspace{6pt} 


\authorcontributions{Conceptualization, A.M. and A.S.; methodology, A.M.;  formal analysis, A.M., A.S. and T.M.; investigation, A.M. and A.S.; writing---original draft preparation, A.M; writing---review and editing, A.M., A.S. and T.M. All authors have read and agreed to the published version of the manuscript.}

\funding{ 
The work was performed as part of the SAO RAS government contract approved by the Ministry of Science and Higher Education of the Russian Federation.
}


\acknowledgments{We obtained  the observed data on the unique scientific facility ``Big Telescope Alt-azimuthal''  of SAO RAS.  The renovation of telescope equipment is currently provided within the national project ``Science and Universities''.  We thank Roman Uklein and Dmitry Oparin who performed  observations at the 2.5m and 6m telescopes.
Some of the data presented in this paper were obtained from the Mikulski Archive for Space Telescopes (MAST). This research has made  use of the NASA/IPAC Extragalactic Database (NED), which is funded by the National Aeronautics and Space Administration and operated by the California Institute of Technology.

The work used  the public data of the Legacy Surveys (http://legacysurvey.org), that  consists of three individual and complementary projects: the Dark Energy Camera Legacy Survey (DECaLS; Proposal ID $\sharp 2014B-0404$;
PIs: David Schlegel and Arjun Dey), the Beijing-Arizona Sky Survey (BASS; NOAO Prop. ID $\sharp 2015A-0801$; PIs: Zhou Xu and Xiaohui Fan), and
the Mayall z-band Legacy Survey (MzLS; Prop. ID $\sharp 2016A-0453$; PI: Arjun Dey). DECaLS, BASS and MzLS together include data obtained,
respectively, at the Blanco telescope, Cerro Tololo Inter-American Observatory, NSF’s NOIRLab; the Bok telescope, Steward Observatory,
University of Arizona; and the Mayall telescope, Kitt Peak National Observatory, NOIRLab. The Legacy Surveys project is honored to be permitted
to conduct astronomical research on Iolkam Du\'ag (Kitt Peak), a mountain with particular significance to the Tohono O\'odham Nation.

This project used data obtained with the Dark Energy Camera (DECam), which was constructed by the Dark Energy Survey (DES) collaboration. Funding for the DES Projects has been provided by the U.S. Department of Energy, the U.S. National Science Foundation, the Ministry of Science and Education of Spain, the Science and Technology Facilities Council of the United Kingdom, the Higher Education Funding Council for England, the National Center for Supercomputing Applications at the University of Illinois at Urbana-Champaign, the Kavli Institute of Cosmological Physics at the University of Chicago, Center for Cosmology and Astro-Particle Physics at the Ohio State University, the Mitchell Institute for Fundamental Physics and Astronomy at Texas A\&M University, Financiadora de Estudos e Projetos, Fundacao Carlos Chagas Filho de Amparo, Financiadora de Estudos e Projetos, Fundacao Carlos Chagas Filho de Amparo a Pesquisa do Estado do Rio de Janeiro, Conselho Nacional de Desenvolvimento Cientifico e Tecnologico and the Ministerio da Ciencia, Tecnologia e Inovacao, the Deutsche Forschungsgemeinschaft and the Collaborating Institutions in the Dark Energy Survey. The Collaborating Institutions are Argonne National Laboratory, the University of California at Santa Cruz, the University of Cambridge, Centro de Investigaciones Energeticas, Medioambientales y Tecnologicas-Madrid, the University of Chicago, University College London, the DES-Brazil Consortium, the University of Edinburgh, the Eidgenossische Technische Hochschule (ETH) Zurich, Fermi National Accelerator Laboratory, the University of Illinois at Urbana-Champaign, the Institut de Ciencies de l’Espai (IEEC/CSIC), the Institut de Fisica de Altes Energies, Lawrence Berkeley National Laboratory, the Ludwig Maximilians Universitat Munchen and the associated Excellence Cluster Universe, the University of Michigan, NSF’s NOIRLab, the University of Nottingham, the Ohio State University, the University of Pennsylvania, the University of Portsmouth, SLAC National Accelerator Laboratory, Stanford University, the University of Sussex, and Texas A\&M University.

BASS is a key project of the Telescope Access Program (TAP), which has been funded by the National Astronomical Observatories of China, the Chinese Academy of Sciences (the Strategic Priority Research Program ‘The Emergence of Cosmological Structures’ Grant \#XDB09000000), and the Special Fund for Astronomy from the Ministry of Finance. The BASS is also supported by the External Cooperation Program of Chinese Academy of Sciences (Grant \# 114A11KYSB20160057), and Chinese National Natural Science Foundation (Grant \# 11433005).

The Legacy Survey team makes use of data products from the Near-Earth Object Wide-field Infrared Survey Explorer (NEOWISE), which is a project of the Jet Propulsion Laboratory/California Institute of Technology. NEOWISE is funded by the National Aeronautics and Space Administration.

The Legacy Surveys imaging of the DESI footprint is supported by the Director, Office of Science, Office of High Energy Physics of the U.S. Department of Energy under Contract No. DE-AC02-05CH1123; by the National Energy Research Scientific Computing Center, a DOE Office of Science User Facility under the same contract; and by the U.S. National Science Foundation, Division of Astronomical Sciences under Contract No. AST-0950945 to NOAO.
}

\conflictsofinterest{The authors declare no conflict of interest.} 

\abbreviations{Abbreviations}{
The following abbreviations are used in this manuscript:\\
\noindent 
\begin{tabular}{@{}ll}
AGN  & Active galaxy nucleus\\
EELR & Extended emission-line region\\
FWHM & Full width at half-maximum\\
LINER & Low-ionization nuclear emission-line region \\
SAO RAS & Special Astrophysical Observatory of the Russian Academy of Sciences \\
\end{tabular}
}

\begin{adjustwidth}{-\extralength}{0cm}

\reftitle{References}

\end{adjustwidth}
\end{document}